\documentclass
[superscriptaddress,secnumarabic,amssymb,amsmath,nobibnotes,aps,prd,showkeys,showpacs,nofootinbib,onecolumn]{revtex4}%
\usepackage{graphicx}
\usepackage{epsf}
\usepackage{bm}
\usepackage{amsmath}
\usepackage{amsfonts}
\usepackage{amssymb}
\usepackage{color}%
\providecommand{\U}[1]{\protect\rule{.1in}{.1in}}

\newcommand{\be}{\begin{equation}}
\newcommand{\ee}{\end{equation}}

\newcommand{\mincir}{\raise
-3.truept\hbox{\rlap{\hbox{$\sim$}}\raise4.truept\hbox{$<$}\ }}
\newcommand{\magcir}{\raise
-3.truept\hbox{\rlap{\hbox{$\sim$}}\raise4.truept\hbox{$>$}\ }}

\begin{document}
\title{Exact Solutions in Chiral Cosmology}
\author{Andronikos Paliathanasis}
\email{anpaliat@phys.uoa.gr}
\affiliation{Institute of Systems Science, Durban University of Technology, PO Box 1334,
Durban 4000, Republic of South Africa}
\author{Genly Leon}
\email{genly.leon@ucn.cl}
\affiliation{Departamento de Matem\'{a}ticas, Universidad Cat\'{o}lica del Norte, Avda.
Angamos 0610, Casilla 1280 Antofagasta, Chile}
\author{Supriya Pan}
\email{supriya.maths@presiuniv.ac.in}
\affiliation{Department of Mathematics, Presidency University, 86/1 College Street, Kolkata
700073, India}

\begin{abstract}
In multi-scalar field cosmologies new dynamical degrees of freedom are
introduced which can explain the observational phenomena. Unlike the usual
scalar field theory where a single scalar field is considered, the
multi-scalar field cosmologies allow more than one scalar field and exhibits
interetsing consequences, such as quintom, hybrid inflation etc. The current
work study the existence of exact solutions and integrable dynamical systems
in multi-scalar field cosmology and more specifically in the so-called Chiral
cosmology where nonlinear terms exists in the kinetic term of the scalar
fields. We present the exact analytic solutions for a system of $N$-scalar
fields. In particular, we consider a multi scalar field cosmological scenario
comprised of $N$-scalar fields that are minimally coupled to the Einstein
gravity. The geometry of the universe is described by the spatially flat
homogeneous and isotropic line element and the scalar fields may interact in
their kinetic or/and potential terms. Within this set up, we show that for a
specific geometry in the kinetic part of the scalar fields and specific
potential form, the gravitational field equations for the class of $N$-scalar
field models can be exactly solved. More specifically, we show that the
Einstein field equations in $N-$scalar field cosmology can be reduced to that
of a $\left(  N+1\right)  $-linear system.

\end{abstract}
\maketitle
\date{\today}

\section{Introduction}

Scalar fields play a significant role in cosmological studies because they can
describe various phenomena of our universe such as the inflationary era
(inflaton), the late time acceleration (dark energy), the dark matter
component in the universe, unification of early inflation to late acceleration
\cite{lid02,ref1a,ref1,newinf,ref4,peebles,tsuwi,lid011,vs,tmatos,urena1,Peebles:1998qn,deHaro:2016hpl,deHaro:2016cdm}%
. Scalar fields are defined to be either minimally or non-minimally coupled to
the gravity \cite{ratra01,bran01,ohanlon,hor,gali} while they also can
attribute, in a dynamical way, the higher-order derivatives of modified
theories of gravity, the later is achieved with the use of Lagrange
multipliers \cite{lan001,lan002,lan003,lan004,lan005}. Moreover, scalar fields
also play an essential role in the conformal transformation which relates the
Einstein and the Jordan frames \cite{ej00}; hence their study is important in
order to understand the physical properties of conformal transformations and
relation of solutions between the different frames \cite{ej01,ej02,ej03}.

In the present work we concentrate on a specific cosmological scenario which
has drawn a remarkable attention to the scientific society in last several
years, is the multi-scalar field theory. In particular, two or more scalar
fields which in general interact in the potential or/and in the kinetic terms,
have been considered earlier to describe the evolution of the universe. The
multi-scalar field models provide new degrees of freedom which can be used to
connect with the phenomena described in string theory to cosmology and
phenomenology \cite{str00,str1}. Furthermore, the multi-scalar fields can also
be defined in the Einstein and in the Jordan frames. In the later scenario,
they describe tensor multi-scalar field models \cite{giampero}. An interesting
connection of the tensor multi-scalar field models is that they are related
with multidimensional cosmologies \cite{multi1}.

The simplest multi-scalar field theory is the quintom theory in which the two
scalar fields, namely a quintessence scalar field and a phantom field
contribute to the dark sector, specifically the dark energy, of the universe
\cite{quin00,quin2,quin3,quin4,quin5}. One of the main properties of such
quintom models is that the equation-of-state parameter for the dark energy
fluid is able to evolve across the cosmological constant boundary `$-1$'
\cite{quin00,od1}. Moreover, multi-scalar field models are considered for the
description of inflation \cite{hy1,hy2,hy3}, such as hybrid inflation, double
inflation, $\alpha$-attractors etc., as these models can provide a different
exit from the standard inflationary era \cite{hy4,hy5,hy6,atr1,atr2,atr3}.
Hybrid inflation is \cite{hy1} described by a two-scalar fields model where
the one field describes the inflaton and the second field is an
auxiliary\ Higgs-type field. In such a inflationary model the exit from the
inflation is described by a waterfall regime where the second field is rapid
rolling when inflaton reaches a specific value \cite{water1}. \ Other
cosmological applications of multi-scalar field models with or without
interaction can be found in \cite{re1,re2,re3,re4,re5,re6,re7,re8,re9} and
references therein. The multi-scalar fields can describe different states for
the matter source of the universe; for instance, the multi-scalar fields can
describe the dark sector, i.e., dark energy and dark matter, while the
interaction terms between the scalar fields provide an interaction between the
different fluid terms. In terms of the phenomenology, this kind of
interactions provide viable cosmological parameters \cite{exp1,exp2,exp3}.

Concerning analytic solutions of the gravitational field equations, there are
various analytic and exact solutions in the literature on scalar field
cosmology in homogeneous spacetimes. Most of the solutions correspond to a
Friedmann--Lema\^{\i}tre--Robertson--Walker (FLRW) geometry
\cite{fr01,fr02,fr03,fr04,fr05,fr06,fr07,fr08,fr09} while only a few analytic
solutions are known for anisotropic Bianchi spacetimes \cite{bia1,bia2}. In
the case of multi-scalar field scenarios, there are a small number of exact
and analytic solutions known in the literature with or without the interaction
between the scalar fields
\cite{ChimentoTwoSF,Moraes,Bazeia,arefeva,ants,chir1,chir3,no1,kess,ss1}.

In this work we discuss a family of exact solutions for $N-$scalar field
cosmologies where the scalar fields interact in the kinetic and in the
potential terms, the so-called Chiral cosmological model. In particular, we
have seen that the scalar fields are defined in a hyperbolic geometry, i.e.,
in a space of negative constant curvature, and when $N=2$ they reduce to the
sigma models \cite{sigm0}, which have various applications in inflation
\cite{chir3} as well as in the late-time accelerating phase of the universe
\cite{sigm1}.\ The plan of the paper is the following.

In Section \ref{sec2}, we present the gravitational field equations of multi
scalar field models in a flat FLRW geometry. In Section \ref{sec3} we describe
the exact analytical solutions for such multi scalar field cosmologies.
Finally, in Section \ref{conclu} we conclude the present work with a brief summary.

\section{Chiral Cosmology}

\label{sec2}

The recent observations support the cosmological principle that means in large
scales our universe is isotropic and homogeneous \cite{kalus1}. Thus, in the
context of General Relativity the natural spacetime is described by the FLRW
geometry. Moreover, the FLRW universe with zero spatial curvature is
preferable by the recent observational evidences \cite{planck}, which leads
the spacetime to be described by the following line element%
\begin{equation}
ds^{2}=-dt^{2}+a^{2}\left(  t\right)  \left(  dx^{2}+dy^{2}+dz^{2}\right)  .
\label{msf.01}%
\end{equation}
where $a(t)$ is the expansion scale factor of the universe.

We consider the Action Integral%
\begin{equation}
S=\int\sqrt{-g}dx^{4}R+\int\sqrt{-g}dx^{4}L_{\Phi}\left(  \Phi^{C},\nabla
_{\mu}\Phi^{C}\right)  \label{msf.02}%
\end{equation}
where $L_{\Phi}\left(  \Phi^{B},\nabla\Phi^{B}\right)  $ describes the
Lagrangian of the $N$-scalar fields given by the following expression
\cite{alfaro1,alfaro2,alfaro3}.%
\begin{equation}
L_{\Phi}\left(  \Phi^{C},\nabla_{\mu}\Phi^{C}\right)  =-\frac{1}{2}g^{\mu\nu
}J_{AB}\left(  \Phi^{C}\right)  \nabla_{\mu}\Phi^{A}\nabla_{\nu}\Phi
^{B}-V\left(  \Phi^{C}\right)  \label{msf.03}%
\end{equation}
where $\mu,\nu=1,2,3,4$ and $A,B=1,2...N.$ The tensor $g_{\mu\nu}$ denotes the
natural spacetime and $J_{AB}\left(  \Phi^{C}\right)  $ describes the space in
which the $N-$fields are defined. For instance, for $N=2$, quintom model is
recovered when $J_{AB}\left(  \Phi^{C}\right)  =\left(  1,-1\right)  $, while
the $\sigma-$model is recovered when $J_{AB}\left(  \Phi^{C}\right)  =\left(
1,e^{2\Phi_{1}}\right)  $.

At this point we would like to remark that these are not the only choices
where the quintom model or the (special case of) $\sigma$-model are recovered.
In particular, for every second rank tensor $J_{AB}\left(  \Phi^{C}\right)  $
which defines the $M^{2}$ space, quintom model is recovered, but for different
functional forms of the potential function $V\left(  \Phi^{C}\right)  $. On
the other hand, when $J_{AB}\left(  \Phi^{C}\right)  ~$describes a maximally
symmetric space with negative curvature, the $\sigma$-model is recovered.

Now, the variation of the Action Integral (\ref{msf.02}) with respect to the
metric tensor provides the Einstein field equations
\begin{equation}
G_{\mu\nu}=T_{\mu\nu}\left(  \Phi^{C},\nabla_{\mu}\Phi^{C}\right)  ,
\label{msf.04}%
\end{equation}
in which the energy-momentum tensor assumes the following expression
\begin{equation}
T_{\mu\nu}\left(  \Phi^{C},\nabla_{\mu}\Phi^{C}\right)  =J_{AB}\left(
\Phi^{C}\right)  \nabla_{\mu}\Phi^{A}\nabla_{\nu}\Phi^{B}-g_{\mu\nu}\left(
-\frac{1}{2}g^{\mu\nu}J_{AB}\left(  \Phi^{C}\right)  \nabla_{\mu}\Phi
^{A}\nabla_{\nu}\Phi^{B}-V\left(  \Phi^{C}\right)  \right)  , \label{msf.05}%
\end{equation}
while the variation of the Action Integral (\ref{msf.02}) with respect to
$\Phi^{A}$ gives%
\begin{equation}
g^{\mu\nu}\left(  \nabla_{\mu}J_{~B}^{A}\left(  \Phi^{C}\right)  \nabla_{\nu
}\Phi^{B}\right)  +J_{~B}^{A}\left(  \Phi^{C}\right)  \frac{\partial V\left(
\Phi^{C}\right)  }{\partial\Phi^{B}}=0. \label{msf.06}%
\end{equation}

For the line element (\ref{msf.01}) and by assuming that the $N-$fields
inherits the isometries of the FLRW space the field equations (\ref{msf.04}),
(\ref{msf.06}) are simplified as follows%
\begin{equation}
-3H^{2}+\frac{1}{2}J_{AB}\left(  \Phi^{C}\right)  \dot{\Phi}^{A}\dot{\Phi}%
^{B}+V\left(  \Phi^{C}\right)  =0, \label{msf.07}%
\end{equation}%
\begin{equation}
-2\dot{H}-3H^{2}+\frac{1}{2}J_{AB}\left(  \Phi^{C}\right)  \dot{\Phi}^{A}%
\dot{\Phi}^{B}-V\left(  \Phi^{C}\right)  =0, \label{msf.08}%
\end{equation}
and%
\begin{equation}
\ddot{\Phi}^{A}+\tilde{\Gamma}_{BC}^{A}\left(  \Phi^{D}\right)  \dot{\Phi}%
^{B}\dot{\Phi}^{C}+3H\dot{\Phi}^{A}+J^{AB}\left(  \Phi^{C}\right)
V_{,B}\left(  \Phi^{C}\right)  =0. \label{msf.09}%
\end{equation}
where an overhead dot means the derivative with respect the variable
\textquotedblleft$t$\textquotedblright; $H\equiv\dot{a}/a$ is the Hubble
function and $\tilde{\Gamma}_{BC}^{A}\left(  \Phi^{D}\right)  $ denotes the
symmetric connection coefficients of $J_{AB}\left(  \Phi^{C}\right)  $,
defined as%
\begin{equation}
\tilde{\Gamma}_{BC}^{A}\left(  \Phi^{D}\right)  =\frac{1}{2}J^{AD}\left(
J_{AD,C}+J_{DB,C}-J_{AB,D}\right)  . \label{msf.09a}%
\end{equation}

The field equations can be written, in an equivalent way, in terms of the
energy density $\rho_{\Phi}$ and pressure $p_{\Phi}$ as%
\begin{equation}
-3H^{2}+\rho_{\Phi}=0, \label{msf.10}%
\end{equation}%
\begin{equation}
-2\dot{H}-3H^{2}+p_{\Phi}=0, \label{msf.11}%
\end{equation}
and%
\begin{equation}
\dot{\rho}_{\Phi}+3H\left(  \rho_{\Phi}+p_{\Phi}\right)  =0, \label{msf.12}%
\end{equation}
in which the energy density and the pressure of the scalar field, respectively
given by
\begin{equation}
\rho_{\Phi}=\frac{1}{2}J_{AB}\left(  \Phi^{C}\right)  \dot{\Phi}^{A}\dot{\Phi
}^{B}+V\left(  \Phi^{C}\right)  , \label{msf.13}%
\end{equation}%
\begin{equation}
p_{\Phi}=\frac{1}{2}J_{AB}\left(  \Phi^{C}\right)  \dot{\Phi}^{A}\dot{\Phi
}^{B}-V\left(  \Phi^{C}\right)  , \label{msf.14}%
\end{equation}
describe the components of the total fluid source, that is%
\begin{equation}
\rho_{\Phi}=%
{\displaystyle\sum\limits_{A=1}^{N}}
\rho_{A}\left(  \Phi^{C},\nabla_{\mu}\Phi^{C}\right)  ~,~p_{\Phi}=%
{\displaystyle\sum\limits_{A=1}^{N}}
p_{A}\left(  \Phi^{C},\nabla_{\mu}\Phi^{C}\right)  \ \label{msf.15}%
\end{equation}
where the definitions of $\rho_{A}\left(  \Phi^{C},\nabla_{\mu}\Phi
^{C}\right)  ,~p_{A}\left(  \Phi^{C},\nabla_{\mu}\Phi^{C}\right)  ~$depend on
the assumptions for the model of study every time. However, the time evolution
of the total fluid source and of the scalar factor is independently from the
definition of the energy density for the individual fields.

The field equations (\ref{msf.07})-(\ref{msf.09}) are not independent. The
first-order equation (\ref{msf.07}) can be seen as a conservation law for the
second-order equations (\ref{msf.08})-(\ref{msf.09}). That means, the
integration constants of a solution of (\ref{msf.08})-(\ref{msf.09}) are
constrained by eq. (\ref{msf.07}). Moreover, the degrees of freedom of the
field equations (\ref{msf.08})-(\ref{msf.09}) is $\left(  N+1\right)  $, and
because equation (\ref{msf.07}) is a conservation law, thus, one needs to find
the $N$-linear independent integrals of motion, which are in involution, in
order to determine (Liouville-Arnold) integrability. Consequently, to write
the solutions in closed-form expressions.

Thereafter, in order to determine the exact and analytic solutions for the
model of our consideration we should determine the conservation laws. That can
be achieved by many ways, however, two of the most systematic ways are the Lie
theory and the singularity analysis. These two theories provide new
constraints for the determination of the unknown functions of the our model,
namely, $J_{AB}\left(  \Phi^{C}\right)  $ and $V\left(  \Phi^{C}\right)  $,
which lead to integrable systems \cite{ants,lie2,lie3,lie4}.

In contrast to that approaches, in this work we follow the inverse steps.
Specifically, we consider \textquotedblleft classical\textquotedblright%
\ well-known integrable systems and then we determine the equivalent
multi-scalar field models.

\section{Exact Solutions}

\label{sec3}

Cosmological constant $\Lambda$ is the simplest candidate for the dark energy
fluid which drives the accelerating phase of the universe, and it has been
found to be in well agreement with a series of astronomical observations.
While it is well known that $\Lambda$ suffers from two major problems, namely,
the fine-tuning and the coincidence problems \cite{Padmanabhan1,Weinberg1},
but it is the dark energy model with a minimal degrees of freedom. Moreover,
in this context, the evolution of the scale factor is given by an exact
expression, and in particular by an exponential function.

In $\Lambda$-cosmology, assuming the absence of any matter source, the field
equations are
\begin{equation}
-3a\dot{a}^{2}+2a^{3}\Lambda=0 \label{msf.16}%
\end{equation}
and
\begin{equation}
\ddot{a}+\frac{1}{2a}\dot{a}^{2}-a\Lambda=0. \label{msf.17}%
\end{equation}

The latter system describes the de Sitter universe where $a\left(  t\right)
=a_{0}\exp\left(  \sqrt{\frac{2\Lambda}{3}}\;t\right)  $. While expression
(\ref{msf.16})-(\ref{msf.17}) is a set of nonlinear differential equations
that can be linearized under the change of variables $a\left(  t\right)
=r\left(  t\right)  ^{\frac{2}{3}}$. Indeed they can be written in the
equivalent form as follows \cite{ahoj}
\begin{equation}
-\frac{1}{2}\dot{r}^{2}+\frac{\omega^{2}}{2}r^{2}=0~,~~\ddot{r}-\omega
^{2}r=0~~\text{with}~~\omega^{2}=\frac{3}{2}\Lambda. \label{msf.18}%
\end{equation}

These linear equations form a maximally symmetric system \cite{lutz} which is
nothing else than the one-dimensional \textquotedblleft
oscillator\textquotedblright\ (hyperbolic for $\Lambda>0$, harmonic for
$\Lambda<0$).

\subsection{$1+1$ degrees of freedom}

The quintessence UDM (unified dark matter) cosmological models
\cite{Ber07,Gorini04,Gorini05,BasilLukes} provide a de Sitter point in the
evolution of the universe as a late-time attractor while they can also provide
a component in the Hubble function similar to that of the pressureless fluid
term. In Ref. \cite{BasilLukes}, it has been shown that the UDM\ scalar field
model is in a fair agreement with that of the $\Lambda$-cosmology at the
background and perturbation levels, while some recent cosmological constraints
show that the UDM model as the dark energy candidate is observational
supported \cite{fr05}.

Except from the above physical properties, the UDM scalar field model has an
important mathematical property $-$ it is integrable as is described by the
Liouville-Arnold theorem. The potential function of the UDM model is given by
the expression%
\begin{equation}
V\left(  \phi\right)  =\left(  V_{0}+V_{1}\cosh^{2}\left(  \sqrt{\frac{3}{8}%
}\phi\right)  \right)  , \label{msf.19}%
\end{equation}
while the field equations can be written in a linear form under the point
transformation \cite{fr05}%
\begin{equation}
a^{3}=\frac{3}{8}\left(  x^{2}-y^{2}~\right)  ,~\phi=\sqrt{\frac{8}{3}}\arctan
h\left(  \frac{y}{x}\right)  , \label{msf.20}%
\end{equation}
as follows%
\begin{equation}
-\frac{1}{2}\dot{x}^{2}+\frac{1}{2}\dot{y}^{2}+\frac{\omega_{1}^{2}}{2}%
x^{2}-\frac{\omega_{2}^{2}}{2}y^{2}=0, \label{msf.21}%
\end{equation}
\qquad%
\begin{equation}
\ddot{x}-\omega_{1}^{2}x=0~,~\ddot{y}-\omega_{2}^{2}y=0 \label{msf.22}%
\end{equation}
in which $\omega_{1}^{2}=\omega_{1}^{2}\left(  V_{0},V_{1}\right)  $~and
$\omega_{2}^{2}=\omega_{1}^{2}\left(  V_{0},V_{1}\right)  $. At this point it
is important to remark that the same property holds either when $\phi$ is a
phantom scalar field\footnote{When $\phi$ is a phantom field, the
corresponding UDM potential is $V\left(  \phi\right)  =\left(  V_{0}+V_{1}%
\cos^{2}\sqrt{\frac{3}{8}}\phi\right)  $ and the point transformation which
linearizes the field equations is defined as $a^{3}=\frac{3}{8}\left(  \bar
{x}^{2}+\bar{y}^{2}~\right)  ,~\phi=\sqrt{\frac{8}{3}}\arctan\left(
\frac{\bar{y}}{\bar{x}}\right)  $.}.

However, the most generic 1+1 dimensional Lagrangian which describes a linear
system is
\begin{equation}
L\left(  x,\dot{x},y,\dot{y}\right)  =-\frac{1}{2}\dot{x}^{2}+\frac{1}{2}%
\dot{y}^{2}-\frac{\omega_{1}^{2}}{2}x^{2}+\frac{\omega_{2}^{2}}{2}y^{2}%
-\mu^{2}xy, \label{msf.23}%
\end{equation}
with Euler-Lagrange equations%
\begin{equation}
\ddot{x}-\omega_{1}^{2}x-\mu^{2}y=0~,~\ddot{y}-\omega_{2}^{2}y+\mu^{2}x=0,
\label{msf.24}%
\end{equation}
where also we consider the constraint
\begin{equation}
-\frac{1}{2}\dot{x}^{2}+\frac{1}{2}\dot{y}^{2}+\frac{\omega_{1}^{2}}{2}%
x^{2}-\frac{\omega_{2}^{2}}{2}y^{2}+\mu^{2}xy=0. \label{msf.25}%
\end{equation}

Under the inverse point transformation (\ref{msf.20}), the Lagrangian
(\ref{msf.23}) becomes%
\begin{equation}
L\left(  a,\dot{a},\phi,\dot{\phi}\right)  =-3a\dot{a}^{2}+\frac{1}{2}%
a^{3}\dot{\phi}^{2}-\frac{4}{3}a^{3}\left[  \omega_{2}^{2}+\left(  \omega
_{1}^{2}-\omega_{2}^{2}\right)  \cosh^{2}\left(  \sqrt{\frac{3}{8}}%
\phi\right)  +\mu^{2}\sinh\left(  2\sqrt{\frac{3}{8}}\phi\right)  \right]  .
\label{msf.26}%
\end{equation}

The latter potential reduces to the{ UDM model when }${\mu}${$=0$.} Hence, we
can see that by considering the Lagrangian which defines the most general
linear system of second-order differential equations, we are able to construct
a new integrable scalar field model.

As far as concerns the generic solution of (\ref{msf.24}) that is given with
the use of the exponential matrix. Indeed system (\ref{msf.24}) can be written
as%
\begin{equation}%
\begin{pmatrix}
\dot{x}\\
\dot{y}\\
\dot{p}_{x}\\
\dot{p}_{y}%
\end{pmatrix}
=%
\begin{pmatrix}
0 & 0 & 1 & 0\\
0 & 0 & 0 & 1\\
\omega_{1}^{2} & \mu^{2} & 0 & 0\\
-\mu^{2} & \omega_{2}^{2} & 0 & 0
\end{pmatrix}%
\begin{pmatrix}
x\\
y\\
p_{x}\\
p_{y}%
\end{pmatrix}
\label{msf.27}%
\end{equation}
whose general solution is, $\mathbf{x}=\mathbf{x}_{0}e^{\mathbf{A}t}$, where
$\mathbf{A}$ is the $4\times4$ matrix of the latter system and $\mathbf{x}%
_{0}$ is the matrix $\mathbf{x}$ at $t=0$. Hence, the Jordan representation of
matrix $\mathbf{A}$ defines the generic solution of the field equations,
recall that the Jordan representation depends on the free parameters
$\omega_{1},~\omega_{2}$ and $\mu^{2}$.

Now in the simplest case where $\omega_{1}=\omega_{2}=0$, the generic solution
of system (\ref{msf.24}) is
\begin{align}
x\left(  t\right)   &  =\left(  c_{1}\cos\left(  \omega t\right)  +c_{2}%
\sin\left(  \omega t\right)  \right)  \cosh\left(  \omega t\right)  +\left(
c_{3}\cos\left(  \omega t\right)  +c_{4}\sin\left(  \omega t\right)  \right)
\sinh\left(  \omega t\right)  ,\label{msf.28}\\
y\left(  t\right)   &  =\left(  c_{4}\cos\left(  \omega t\right)  -c_{3}%
\sin\left(  \omega t\right)  \right)  \cosh\left(  \omega t\right)  +\left(
c_{2}\cos\left(  \omega t\right)  -c_{1}\sin\left(  \omega t\right)  \right)
\sinh\left(  \omega t\right)  . \label{msf.29}%
\end{align}
where the constraint equation (\ref{msf.26}) provides the algebraic equation
$c_{1}c_{4}-c_{2}c_{3}=0$ and $\omega=\sqrt{2}\mu^{2}$. Consequently, the
latter solution can provide a bouncing universe. For instance, when all the
coefficient constants are equal the scale factor becomes%
\begin{equation}
a^{3}\left(  t\right)  =\frac{3}{4}c_{1}^{2}e^{2\omega t}\sin\left(  2\omega
t\right)
\end{equation}
which provides a periodic universe. Moreover, when $c_{1}=c_{2}$ \ then the
scale factor becomes
\begin{equation}
a^{3}\left(  t\right)  =\frac{3}{8}\left(  c_{2}^{2}-c_{3}^{2}\right)
+\frac{3}{8}\sin\left(  2\omega t\right)  \left(  \left(  c_{2}^{2}+c_{3}%
^{2}\right)  \cosh\left(  2\omega t\right)  +2c_{2}c_{3}\sinh\left(  2\omega
t\right)  \right)
\end{equation}
which is a periodic solution around a constant scale factor, the latter
solution is physically accepted when $c_{2}^{2}-c_{3}^{2}>0$.

\subsection{$2+1$ degrees of freedom}

Let us consider the Lagrangian function%
\begin{equation}
L\left(  x,\dot{x},y,\dot{y},z,\dot{z}\right)  =-\frac{1}{2}\dot{x}^{2}%
+\frac{1}{2}\dot{y}^{2}-\frac{\omega_{1}^{2}}{2}x^{2}+\frac{\omega_{2}^{2}}%
{2}y^{2}+\frac{\omega_{3}^{2}}{2}z^{2}-\mu_{1}^{2}xy-\mu_{2}^{2}xz+\mu_{3}%
^{2}yz, \label{msf.30}%
\end{equation}
which describes the linear system%
\begin{align}
\ddot{x}-\omega_{1}^{2}x-\mu_{1}^{2}y-\mu_{2}^{2}z  &  =0,\label{msf.31}\\
\ddot{y}+\mu_{1}^{2}x-\omega_{2}^{2}y-\mu_{3}^{2}z  &  =0,\label{msf.32}\\
\ddot{z}+\mu_{2}^{2}x-\omega_{3}^{2}z-\mu_{3}^{2}y  &  =0, \label{msf.33}%
\end{align}
with constraint equation the conservation law of \textquotedblleft
energy\textquotedblright\ for the latter system to be zero.

The generic solution of the latter system is given with the us of the
exponential matrix, $\mathbf{x}=\mathbf{x}_{0}e^{\mathbf{A}t}$, in which
$\mathbf{x}=\left(  x,y,z,p_{x},p_{y},p_{z}\right)  ^{T}$ and
\begin{equation}
A=%
\begin{pmatrix}
0 & 0 & 0 & 1 & 0 & 0\\
0 & 0 & 0 & 0 & 1 & 0\\
0 & 0 & 0 & 0 & 0 & 1\\
\omega_{1}^{2} & \mu_{1}^{2} & \mu_{2}^{2} & 0 & 0 & 0\\
-\mu_{1}^{2} & \omega_{2}^{2} & \mu_{3}^{2} & 0 & 0 & 0\\
-\mu_{2}^{2} & \mu_{3}^{2} & \omega_{3}^{3} & 0 & 0 & 0
\end{pmatrix}
. \label{msf.40}%
\end{equation}

Under the point transformation%
\begin{align*}
x  &  =\sqrt{\frac{8}{3}}a^{\frac{3}{2}}\cosh\left(  \sqrt{\frac{3}{8}}%
\phi\right)  ~,\\
y  &  =\sqrt{\frac{8}{3}}a^{\frac{3}{2}}\sinh\left(  \sqrt{\frac{3}{8}}%
\phi~\right)  \cos\left(  \sqrt{\frac{3}{8}}\psi\right)  ~,\\
z  &  =\sqrt{\frac{8}{3}}a^{\frac{3}{2}}\sinh\left(  \sqrt{\frac{3}{8}}%
\phi~\right)  \sin\left(  \sqrt{\frac{3}{8}}\psi\right)  ,
\end{align*}
Lagrangian (\ref{msf.30}) takes the following form%
\begin{align}
L\left(  a,\dot{a},\phi,\dot{\phi},\psi,\dot{\psi}\right)   &  =-3a\dot{a}%
^{2}+\frac{1}{2}a^{3}\left(  \dot{\phi}^{2}+\sinh^{2}\left(  \sqrt{\frac{3}%
{8}}\phi\right)  \dot{\psi}^{2}\right)  +\nonumber\\
&  ~-\frac{4}{3}a^{3}\left[  \omega_{1}^{2}+\left(  \left(  \omega_{1}%
^{2}-\omega_{2}^{2}\right)  +\left(  \omega_{2}^{2}-\omega_{3}^{2}\right)
\sin^{2}\left(  \sqrt{\frac{3}{8}}\psi\right)  -\mu_{3}\sin\left(
2\sqrt{\frac{3}{8}}\psi\right)  \right)  \sinh^{2}\phi\right]  +\nonumber\\
&  -\frac{4}{3}a^{3}\left[  \left(  \mu_{1}\cos\left(  \sqrt{\frac{3}{8}}%
\psi\right)  +\mu_{2}\sin\left(  \sqrt{\frac{3}{8}}\psi\right)  \right)
\sinh\left(  2\sqrt{\frac{3}{8}}\phi\right)  \right]
\end{align}

The latter Lagrangian describes the field equations for a two-scalar fields
model where the two scalar fields interact in the kinetic terms and in the
potential terms as well. In particular, the scalar fields are defining in the
hyperbolic case. The latter model for $\mu_{1}=\mu_{2}=\mu_{3}=0$, has been
found and studied before in \cite{ants} and it is an extension of the UDM
model in the case of two scalar fields.

In a similar fashion we continue with the extension of the exact solutions in
higher-degrees of freedom.

\subsection{$N+1$ degrees of freedom}

In order to generalize our results in case of $N$-scalar fields we assume the
Lagrangian%
\begin{equation}
L\left(  x,\dot{x},y_{\beta},\dot{y}_{\beta}\right)  =-\frac{1}{2}\dot{x}%
^{2}+\frac{1}{2}%
{\displaystyle\sum\limits_{\beta=1}^{N}}
\left(  \dot{y}_{\beta}\right)  ^{2}-\frac{\omega_{0}^{2}}{2}x^{2}+\frac{1}{2}%
{\displaystyle\sum\limits_{\beta=1}^{N}}
\left(  \omega_{\beta}y_{\beta}\right)  ^{2}-%
{\displaystyle\sum\limits_{\beta=1}^{N}}
\left(  \mu_{\beta}^{2}xy_{\beta}\right)  +%
{\displaystyle\sum\limits_{\left(  \beta\neq\gamma\right)  \beta,\gamma=1}%
^{N}}
\left(  \mu_{\beta\gamma}^{2}y_{\beta}y_{\gamma}\right)  \label{msf.41}%
\end{equation}
which describes the following system of linear equations%
\begin{equation}
\ddot{x}-\omega_{0}^{2}x-%
{\displaystyle\sum\limits_{\beta=1}^{N}}
\left(  \mu_{\beta}^{2}y_{\beta}\right)  =0, \label{msf.42}%
\end{equation}%
\begin{equation}
\ddot{y}_{\beta}+\mu_{\beta}^{2}x-\omega_{\beta}^{2}y_{\beta}-%
{\displaystyle\sum\limits_{\left(  \beta\neq\gamma\right)  ~\gamma=1}^{N}}
\left(  \mu_{\beta\gamma}^{2}y_{\gamma}\right)  =0. \label{msf.43}%
\end{equation}
Moreover, we consider the constraint condition%
\begin{equation}
-\frac{1}{2}\dot{x}^{2}+\frac{1}{2}%
{\displaystyle\sum\limits_{\beta=1}^{N}}
\left(  \dot{y}_{\beta}\right)  ^{2}+\frac{\omega_{0}^{2}}{2}x^{2}-\frac{1}{2}%
{\displaystyle\sum\limits_{\beta=1}^{N}}
\left(  \omega_{\beta}y_{\beta}\right)  ^{2}+%
{\displaystyle\sum\limits_{\beta=1}^{N}}
\left(  \mu_{\beta}^{2}xy_{\beta}\right)  -%
{\displaystyle\sum\limits_{\left(  \beta\neq\gamma\right)  \beta,\gamma=1}%
^{N}}
\left(  \mu_{\beta\gamma}^{2}y_{\beta}y_{\gamma}\right)  =0. \label{msf.44}%
\end{equation}
The latter condition is important in order the generic solution of
(\ref{msf.42})-(\ref{msf.43}) to be valid for the case of multi-scalar fields model.

Hence, system (\ref{msf.42})-(\ref{msf.43}) can be written in the equivalent
form
\begin{equation}
\mathbf{\dot{X}}=\mathbf{A~X,} \label{msf.45}%
\end{equation}
$~\ \ \ \ \ $with $X=\left(  x,y_{\beta},p_{x},p_{\beta}\right)  ^{T}$, and
\begin{equation}
\mathbf{A}=%
\begin{pmatrix}
0 & \mathbf{I}\\
\mathbf{B} & 0
\end{pmatrix}
~, \label{msf.46}%
\end{equation}
in which $\mathbf{I}$ is the unitary matrix of dimension $\left(  N+1\right)
$ and$B$ is an $\left(  N+1\right)  \times\left(  N+1\right)  $ matrix defined
as
\begin{equation}
\mathbf{B}=%
\begin{pmatrix}
\omega_{0}^{2} & \mu_{1} & \mu_{2} & ... & \mu_{N}\\
-\mu_{1} & \omega_{1}^{2} & \mu_{12} & ... & \mu_{1N}\\
-\mu_{2} & \mu_{21} & \omega_{2}^{2} & ... & \mu_{2N}\\
... & ... & ... & ... & ...\\
-\mu_{N} & \mu_{N1} & ... & \mu_{N2} & \omega_{N}^{2}%
\end{pmatrix}
. \label{msf.47}%
\end{equation}
Thus, the functional form of the solution of system (\ref{msf.45}) depends on
the eigenvalues of the matrix $\mathbf{B}$, and consequently on the values of
the free parameters $\mathbf{\omega}$ and $\mathbf{\mu}$.

We consider the point transformation%
\begin{align*}
x  &  =\sqrt{\frac{8}{3}}a^{\frac{3}{2}}\cosh\left(  \sqrt{\frac{3}{8}}%
\phi_{1}\right) \\
y_{1}  &  =\sqrt{\frac{8}{3}}a^{\frac{3}{2}}\sinh\left(  \sqrt{\frac{3}{8}%
}\phi_{1}\right)  \cos\left(  \sqrt{\frac{3}{8}}\phi_{2}\right) \\
y_{2}  &  =\sqrt{\frac{8}{3}}a^{\frac{3}{2}}\sinh\left(  \sqrt{\frac{3}{8}%
}\phi_{1}\right)  \sin\left(  \sqrt{\frac{3}{8}}\phi_{2}\right)  \cos\left(
\sqrt{\frac{3}{8}}\phi_{3}\right) \\
&  ...\\
y_{N-1}  &  =\sqrt{\frac{8}{3}}a^{\frac{3}{2}}\sinh\left(  \sqrt{\frac{3}{8}%
}\phi_{1}\right)
{\displaystyle\prod\limits_{\beta=2}^{N-1}}
\sin\left(  \sqrt{\frac{3}{8}}\phi_{\beta}\right)  ~\cos\left(  \sqrt{\frac
{3}{8}}\phi_{N}\right) \\
y_{N}  &  =\sqrt{\frac{8}{3}}a^{\frac{3}{2}}\sinh\left(  \sqrt{\frac{3}{8}%
}\phi_{1}\right)
{\displaystyle\prod\limits_{\beta=2}^{N}}
\sin\left(  \sqrt{\frac{3}{8}}\phi_{\beta}\right)
\end{align*}
where the Lagrangian (\ref{msf.41}) takes the following form%
\begin{align}
L\left(  a,\dot{a},\phi_{\beta},\dot{\phi}_{\beta}\right)   &  =-3aa^{2}%
+\frac{1}{2}a^{3}\left[  \left(  \dot{\phi}_{1}\right)  ^{2}+\sinh^{2}\left(
\sqrt{\frac{3}{8}}\phi_{1}\right)  \left(  \dot{\phi}_{2}\right)  ^{2}%
+\sin^{2}\left(  \sqrt{\frac{3}{8}}\phi_{2}\right)  \left(  \dot{\phi}%
_{3}\right)  ^{2}+%
{\displaystyle\sum\limits_{\beta=3}^{N}}
\left(  \left(  \dot{\phi}_{\beta}\right)  ^{2}%
{\displaystyle\prod\limits_{\gamma=2}^{\beta-1}}
\sin^{2}\left(  \sqrt{\frac{3}{8}}\phi_{\gamma}\right)  \right)  \right]
+\nonumber\\
&  -\frac{4}{3}a^{3}\left(  \omega_{0}^{2}+\left(  \omega_{0}^{2}-\omega
_{1}^{2}\right)  \sinh^{2}\left(  \sqrt{\frac{3}{8}}\phi_{1}\right)  \right)
\nonumber\\
&  -\frac{4}{3}a^{3}\left[  \left(  \left(  \omega_{1}^{2}-\omega_{2}%
^{2}\right)  -\left(  \left(  \omega_{2}^{2}+\omega_{3}^{2}\right)  +\left(
\omega_{3}^{2}+...\right)  \right)  \sin^{2}\left(  \sqrt{\frac{3}{8}}\phi
_{4}\right)  \right)  \sin^{2}\left(  \sqrt{\frac{3}{8}}\phi_{2}\right)
\right] \nonumber\\
&  -\frac{4}{3}a^{3}\sinh\left(  2\sqrt{\frac{3}{8}}\phi_{1}\right)  \left(
{\displaystyle\sum\limits_{\gamma=2}^{N-1}}
\left(
{\displaystyle\prod\limits_{\beta=2}^{\gamma}}
\sin\left(  \sqrt{\frac{3}{8}}\phi_{\beta}\right)  ~\cos\left(  \sqrt{\frac
{3}{8}}\phi_{N}\right)  \right)  +\sinh\left(  \sqrt{\frac{3}{8}}\phi
_{1}\right)
{\displaystyle\prod\limits_{\beta=2}^{N}}
\sin\left(  \sqrt{\frac{3}{8}}\phi_{\beta}\right)  \right) \nonumber\\
&  -\frac{4}{3}a^{3}\sinh\left(  2\sqrt{\frac{3}{8}}\phi_{1}\right)  \left(
{\displaystyle\sum\limits_{\left(  \beta\neq\gamma\right)  \beta,\gamma=2}%
^{N}}
\mu_{\beta}\left(  y_{\beta}y_{\gamma}\right)  \right)  . \label{lan01}%
\end{align}

The above Lagrangian stands for the multi-scalar fields cosmology which is
equivalent to the linear system (\ref{msf.45}). According to our knowledge,
probably this is the first analytic solution for n- interacting scalar field
cosmologies presented in the literature.

\section{Conclusions}

\label{conclu}

The theory of scalar fields has enriched the understanding of the universe
evolution in various ways. From early evolution of the universe to its current
state, scalar field theory has played an essential role
\cite{lid02,ref1a,ref1,newinf,ref4,peebles,tsuwi,lid011,vs,tmatos,urena1,Peebles:1998qn,deHaro:2016hpl,deHaro:2016cdm}%
. In particular, the early inflationary scenario, the intermediate matter
dominated era, late time accelerating phase as well as the unification of
these two accelerating eras have been widely framed into this picture.
Usually, in scalar field theory, we consider a single field with a potential,
however, the scalar field theory consisting of multiple scalar fields are
equally welcome and have been found to offer some exciting results in the
context of universe evolution. It has been found that a system of two scalar
fields can describe some interesting cosmological consequences, such as the
crossing of the cosmological constant boundary `$-1$' (known as quintom models
\cite{quin00,quin2,quin3,quin4}) which for a single scalar field model is
impossible since the single scalar field models can only describe either the
quintessence or the phantom regime. Moreover, these multi-fluid models are
also able to explain the early inflationary universe (known as hybrid
inflation) \cite{hy1,hy2,hy3} with a different graceful exit compared to the
standard inflationary paradigm \cite{hy4,hy5,hy6}. Thus, in many ways,
multi-scalar field models are potentially enriched.

In the present work, we have investigated the exact solutions for minimally
coupled multi-scalar field models. So far we are concerned with the
literature, this is the first time we have presented the exact solutions for
the multi-scalar field models. Assuming a spatially flat FLRW geometry of the
universe with $N$-number of minimally coupled scalar fields in the context of
Einstein gravity, we show that the gravitational field equations can be
exactly solved where the component scalar fields may interact with one another
via their kinetic and/or potential terms.

The study of the integrability of that cosmological models it is important
when we would like to study them either numerical. We can easily infer that
the solutions are not sensitive on the initial conditions which means the
conclusions which can be made are valid for different initial conditions. On
the other hand, these exact solutions can be used as toy models for the
general study of some families of multi-scalar field cosmologies.

The cosmological evolution of these models, and the application of the exact
solutions for the study of various phases of the universe is a subject of
special study, but exceeds the scopus of this work. Such an analysis will be
published in a forthcoming work.

\begin{acknowledgments}
The authors thank Prof. S.V Chervon for useful comments and suggestions. GL
was funded by Comisi\'{o}n Nacional de Investigaci\'{o}n Cient\'{\i}fica y
Tecnol\'{o}gica (CONICYT) through FONDECYT Iniciaci\'{o}n 11180126. GL thanks
to Department of Mathematics and to Vicerrector\'{\i}a de Investigaci\'{o}n y
Desarrollo Tecnol\'{o}gico at Universidad Cat\'{o}lica del Norte for financial support.
\end{acknowledgments}


\begin{thebibliography}{999}                                                                                              %


\bibitem {lid02}A.D. Linde, Phys. Lett. B \textbf{129}, 177 (1983)

\bibitem {ref1a}A.D. Linde, Phys. Lett. B \textbf{108}, 389 (1982)

\bibitem {ref1}A.D. Linde, Phys. Lett. B \textbf{129}, 177 (1983)

\bibitem {newinf}J.D. Barrow, Phys.\ Rev. D \textbf{48}, 1585 (1993)

\bibitem {ref4}J.D. Barrow and P. Saich, Class. Quantum Grav. \textbf{10}, 279 (1993)

\bibitem {peebles}P.J. Peebles and B. Ratra, Astroph. J. Lett. \textbf{325},
L17 (1988)

\bibitem {tsuwi}S.\ Tsuwikawa, Class.\ Quant. Grav. \textbf{30}, 214003 (2013)

\bibitem {lid011}A.R. Liddle and R.J.\ Scherrer, Phys. Rev. D \textbf{59},
023509 (1999)

\bibitem {vs}V. Sahni and L.M. Wang, Phys.\ Rev. D \textbf{62}, 103517 (2000)

\bibitem {tmatos}T. Matos and L.A. Ure\~{n}a-L\'{o}pez, Phys.\ Rev. D
\textbf{63}, 063506 (2001)

\bibitem {urena1}L.A. Ure\~{n}a-L\'{o}pez, J. Phys. Conf. Series \textbf{761},
012076 (2016)

\bibitem {Peebles:1998qn}P.~J.~E.~Peebles and A.~Vilenkin, Phys.\ Rev.\ D
\textbf{59}, 063505 (1999)

\bibitem {deHaro:2016hpl}J.~de Haro, J.~Amor\'{o}s and S.~Pan, Phys.\ Rev.\ D
\textbf{93}, 084018 (2016)

\bibitem {deHaro:2016cdm}J.~de Haro, J.~Amor\'{o}s and S.~Pan, Phys.\ Rev.\ D
\textbf{94}, 064060 (2016)

\bibitem {ratra01}P.J. Peebles and B. Ratra, Rev. Mod. Phys. \textbf{75}, 559 (2003)

\bibitem {bran01}C.H. Brans and R.H. Dicke, Phys. Rev. \textbf{124}, 925 (1961)

\bibitem {ohanlon}J. O'Hanlon, Phys. Rev. Lett. \textbf{29}, 137 (1972)

\bibitem {hor}G.W. Hordenski, Int. J. Theor. Phys. \textbf{10}, 363 (1974)

\bibitem {gali}C. Deffayet and D.A. Steer, Class. Quant. Grav. \textbf{30},
214006 (2013)

\bibitem {lan001}T.P. Sotiriou, Modifications of Einstein's Theory of Gravity
at Large Distances, eds E. Papantonopoulos, Lecture Notes in Physics, 892,
Springer (2015)

\bibitem {lan002}D. S\'{a}ez-G\'{o}mez, Phys. Rev. D \textbf{85}, 023009 (2012)

\bibitem {lan003}A. Paliathanasis, S. Pan and S. Pramanik, Class. Quant. Grav.
\textbf{32}, 245006 (2015)

\bibitem {lan004}A.\ Paliathanasis, Phys.\ Rev. D \textbf{95}, 064062 (2017)

\bibitem {lan005}J. Kluson, Class. Quant. Grav. \textbf{28}, 125025 (2011)

\bibitem {ej00}V. Faraoni, E. Gunzig and P. Nardone, Fund. Cosm. Phys.
\textbf{20}, 121 (1999)

\bibitem {ej01}V. Faraoni and E. Gunzig, Int. J. Theor. Phys. \textbf{38}, 217 (1999)

\bibitem {ej02}M. Postma and M. Volponi, \ Phys.\ Rev. D \textbf{90}, 103516 (2014)

\bibitem {ej03}N. Dimakis, A. Giacomini and A. Paliathanasis, Eur. Phys. J. C
\textbf{78}, 751 (2018)

\bibitem {str00}A. Collinucci, M. Nielsen and T. Van Riet, Class.\ Quant.
Grav. \textbf{22}, 1269 (2005)

\bibitem {str1}A. Achuracco and G.A. Palma, [arXiv:1807.04390]

\bibitem {giampero}T. Damouri and G. Espsito-Farese, Class.\ Quant. Grav.
\textbf{9}, 2093 (1992)

\bibitem {multi1}M. Rainer and A.\ Zhuk, Phys. Rev. D \textbf{54}, 6186 (1996)

\bibitem {quin00}Y.F. Cai, E.N. Saridakis, M.R.\ Setare and J.-Q. Xia, Phys.
Rep. \textbf{493}, 1 (2010)

\bibitem {quin2}M.R. Setare and E.N. Saridakis, Int. J. Mod. Phys. D
\textbf{18}, 549 (2009)

\bibitem {quin3}R. Lazkoz, G. Leon and I. Quiros, Phys. Lett. B \textbf{649},
103 (2007)

\bibitem {quin4}G. Leon, A. Paliathanasis and J.L. Morales-Martinez, Eur.
Phys. J. C \textbf{78}, 753 (2018)

\bibitem {quin5}E. Elizalde, S. Nojiri, S.D. Odintsov, D. Saez-Gomez and V.
Faraoni, Phys. Rev. D 77, 106005 (2008)

\bibitem {od1}E. Elizalde, S. Nojiri and S.D. Odintsov, Phys. Rev. D 70,
043539 (2004)

\bibitem {hy1}A.D. Lindle, Phys. Rev. D \textbf{49}, 784 (1994)

\bibitem {hy2}E.J. Copeland, A.R. Liddle, D.H. Lyth, E.W. Steward and D.
Wands, Phys. Rev. D \textbf{49}, 6410 (1994)

\bibitem {hy3}S.A. Kim and A.R. Liddle, Phys. Rev. D \textbf{74}, 023513 (2006)

\bibitem {hy4}D. Wands, Lect. Notes Phys. \textbf{738}, 275 (2008)

\bibitem {hy5}J.R. Bond, L. Kofman, S. Prokushkin and P.M. Vaudrevange, Phys.
Rev. D \textbf{75}, 123511 (2007)

\bibitem {hy6}K. Inomata, M. Kawasaki, K. Mukaida and T.T. Yanagida, Phys.
Rev. D \textbf{97}, 043514 (2018)

\bibitem {atr1}P. Carrilho, D. Mulryne, J. Ronaye and T. Tenkanen, JCAP 06,
032 (2018)

\bibitem {atr2}P. Christodoulidis, D. Roest, E.I. Sfakianakis, Angular
inflation in multi-field $\alpha$-attractors, [arXiv:1803.09841]

\bibitem {atr3}P. Christodoulidis, Probing the inflationary evolution using
analytical solutions, [arXiv:1811.06456]

\bibitem {water1}S. Clesse, Phys. Rev.\ D \textbf{83}, 063518 (2011)

\bibitem {water2}D.H.\ Lyth, JCAP \textbf{05}, 022 (2012)

\bibitem {re1}A.A. Coley and R.J. van den Hoogen, Phys. Rev.\ D \textbf{62},
023517 (2000)

\bibitem {re2}S. Tsujikawa, Phys. Rev. D \textbf{73}, 103504 (2006)

\bibitem {re3}Y. Li, Int. J. Mod. Phys. D \textbf{26}, 1750164 (2017)

\bibitem {re4}D.I. Kaiser, Phys. Rev. D \textbf{81}, 084044 (2010)

\bibitem {re5}A.A. Andrianov and O.O. Novikov and C. Lan, Theor. Math. Phys.
\textbf{184}, 1224 (2015)

\bibitem {re6}J.-C. Hwang and H.\ Noh, Phys. Lett. B \textbf{495}, 277 (2000)

\bibitem {re7}P. Cariiho, D. Mulryne, J. Ronaye and T. Tenkanen, JCAP
\textbf{06}, 032 (2018)

\bibitem {re8}A. Aazami and R. Easther, JCAP \textbf{06}, 013 (2006)

\bibitem {re9}J.-O. Gong, Int. J. Mod. Phys. D \textbf{26}, 1740003 (2017)

\bibitem {exp1}A.P. Billyard and A.A. Coley, Phys.\ Rev. D 61, 083503 (2000)

\bibitem {exp2}B.\ Wang, E. Abdalla, F. Atrio-Barandela and D. Pavon, Rept.
Prog. Phys. \textbf{79}, 096901 (2016)

\bibitem {exp3}W. Yang, S. Pan and A. Paliathanasis, Mon. Not. Roy. Astron.
Soc \textbf{482}, 1007 (2019)

\bibitem {fr01}J.G. Russo, Phys. Lett. B \textbf{600}, 185 (2004)

\bibitem {fr02}G.F.R. Ellis and M.S. Madsen, Class. Quant. Grav.
\textbf{8}\ (1991) 667

\bibitem {fr03}H.C. Kim, Mod. Phys. Lett. A \textbf{28}, 135008 (2013)

\bibitem {fr04}C. Rubano and J.D. Barrow, Phys.\ Rev. D \textbf{64}, 127031 (2001)

\bibitem {fr05}A. Paliathanasis, M. Tsamparlis, S. Basilakos and J.D.\ Barrow,
Phys. Rev. D \textbf{91}, 123535 (2015)

\bibitem {fr06}J.D. Barrow and A. Paliathanasis, Phys. Rev. D \textbf{94},
083518 (2016)

\bibitem {fr07}A. Yu. Kamenshchik, E.O. Pozdeeva, A. Tronconi, G. Venturi and
S.Yu. Vernov, Class. Quantum Grav. 31, 105003 (2014)

\bibitem {fr08}A. Yu. Kamenshchik, E.O. Pozdeeva, A. Tronconi, G. Venturi and
S.Yu. Vernov, Class. Quantum Grav. 33, 015004 (2016)

\bibitem {fr09}A.Yu. Kamenshchik, E.O. Pozdeeva, A.A. Starobinsky, A.
Tronconi, G. Venturi and S.Yu. Vernov, Phys. Rev. D 97, 023536 (2018)

\bibitem {bia1}T. Christodoulakis, Th. Grammenos, Ch. Helias, P.G. Kevrekidis
and A. Spanou, J. Math. Phys. \textbf{47}, 042505 (2006)

\bibitem {bia2}M. Tsamparlis and A. Paliathanasis, Gen. Rel.\ Grav.
\textbf{43}, 1861 (2011)

\bibitem {ChimentoTwoSF}L.P. Chimento, Class. Quantum Grav. \textbf{15}, 965 (1998)

\bibitem {Moraes}P.H.R.S. Moraes and J.R.L. Santos, Phys. Rev. D. \textbf{89},
083516 (2014)

\bibitem {Bazeia}D. Bazeia, L. Losano and J.R.L. Santos, Phys. Lett. A
\textbf{377}, 1615 (2013)

\bibitem {arefeva}I.Ya. Aref'eva, N.V. Bulatov and S.Yu. Vernov, Theoretical
and Mathematical Physics, \textbf{163}, 788 (2010)

\bibitem {ants}A. Paliathanasis and M. Tsamparlis, Phys. Rev. D \textbf{90},
43529 (2014)

\bibitem {chir1}S.V Chervon, S.D. Maharaj, A. Beesman and A.S. Kubasov,
Gravit. Cosmol. \textbf{20}, 176 (2014)

\bibitem {chir3}S.V. Chervon, Quantum Matter \textbf{2}, 71 (2013)

\bibitem {no1}Y. Zhang, Y.-G. Gong and Z.-H. Zhu, Phys. Lett. B \textbf{688},
13 (2010)

\bibitem {kess}T. Chiba, A. De Felice and S. Tsujikawa, Phys. Rev. D
\textbf{90}, 023516 (2014)

\bibitem {ss1}A.S.Sakharov and M.Yu.Khlopov, Phys. Atom. Nucl. 56, 412 (1993)

\bibitem {sigm0}S. V. Ketov, Quantum Non-linear Sigma Models, Springer-Verlag,
Berlin, (2000).

\bibitem {sigm1}J. Lee, T.H. Lee, T. Moon and P. Oh, Phys. Rev. D \textbf{80},
0656016 (2009)

\bibitem {kalus1}B. Kalus, D.J. Schwarz, M. Seikel and A. Wiegand, A\&A
\textbf{533}, A56 (2013)

\bibitem {planck}P.A.R. Ade et al. (Planck Collaboration), A\&A \textbf{594},
A13 (2016)

\bibitem {alfaro1}V. de Alfaro, Nuovo Cimento 50A, 523 (1979)

\bibitem {alfaro2}G.G. Ivanov, Theor. Mat. Fiz 57, 45 (1983)

\bibitem {alfaro3}S. Chervon, Russian Physics Journal 38, 5 (1995)

\bibitem {lie2}M. Tsamparlis and A.\ Paliathanasis, Symmetry 10, \textbf{233} (2018)

\bibitem {lie3}A. Paliathanasis, J.D. Barrow and P.G.L. Leach, Phys. Rev.\ D
\textbf{94}, 023525 (2016)

\bibitem {lie4}A. Paliathanasis and P.G.L. Leach, Phys. Lett. A \textbf{380},
2815 (2016)

\bibitem {Padmanabhan1}T. Padmanabhan, Phys. Rept. \textbf{380}, 235 (2003)

\bibitem {Weinberg1}S. Weinberg, Rev. Mod. Phys. \textbf{61}, 1 (1989)

\bibitem {ahoj}A. Paliathanasis, P.G.L. Leach and S. Capozziello, Phys. Lett.
B \textbf{755}, 8 (2016)

\bibitem {lutz}M. Lutzky, J. Phys. A: Math.\ Gen. \textbf{11}, 249 (1978)

\bibitem {Ber07}D. Bertacca, S. Matarrese and M. Pietroni Mod. Phys. Lett. A
\textbf{22} 2893 (2007)

\bibitem {Gorini04}V. Gorini, A. Kamenshchik, U. Moschella, V. Pasquier, Phys.
Rev. D, \textbf{69}, 123512, (2004)

\bibitem {Gorini05}V. Gorini, A. Kamenshchik, U. Moschella, V. Pasquier, A.
Starobinsky, Phys. Rev. D, \textbf{72}, 103518, (2005)

\bibitem {BasilLukes}S. Basilakos and G. Lukes-Gerakopoulos, Phys. Rev. D.
\textbf{78} 083509 (2008)
\end{thebibliography}
\end{document}